# A Resistive CAM Processing-in-Storage Architecture for DNA Sequence Alignment

Roman Kaplan, Leonid Yavits, Ran Ginosar and Uri Weiser
*Dept. of Electrical Engineering, Technion – Israel Institute of Technology*

**Abstract**—A novel processing-in-storage (PRinS) architecture based on Resistive CAM (ReCAM) is described and proposed for Smith-Waterman (S-W) sequence alignment. The ReCAM PRinS massively-parallel compare operation finds matching base-pairs in a fixed number of cycles, regardless of sequence length. The ReCAM PRinS S-W algorithm is simulated and compared to FPGA, Xeon Phi and GPU-based implementations, showing at least 4.7× higher throughput and at least 15× lower power dissipation.

## 1 INTRODUCTION

With the approaching end of Moore's law, academia and industry have an increased interest in non-von Neumann compute paradigms. One example is content addressable associative processing [1]. CMOS-based content addressable memories (CAM) require large bit-cells, limiting chip capacity and forcing most data-intensive applications to employ less functional random access memories. Novel resistive materials dissipate little heat and allow for 3D stacking. Combined with CMOS, resistive materials can be used in a CAM bit-cell, resulting in a small cell area, low leakage power and increased overall chip area efficiency.

This work presents a novel resistive CAM-based storage system architecture with processing-in-storage (PRinS) compute paradigm. The system is an in-storage accelerator that may scale up to hundreds of millions of processing units (PUs) spread across multiple silicon dies, each containing several million PUs. In addition, the system performs the computations in-situ, resulting in increased performance and reduced energy consumption on massively parallel workloads. We name the system Resistive CAM or ReCAM.

The first part of this paper presents ReCAM PRinS system architecture and describes its main components. The second part demonstrates ReCAM PRinS implementation of a key algorithm in bioinformatics, the Smith-Waterman (S-W) DNA local sequence alignment. We also present simulation results and compare the performance of ReCAM PRinS with four state-of-the-art large-scale accelerator systems. We show that an in-storage implementation of S-W on ReCAM may achieve on average 4.7× higher throughput while dissipating 15× lower power compared with a 384-GPU implementation, the largest S-W implementation found in the literature.

The rest of this paper is organized as follows: Section 2 presents the architecture of ReCAM PRinS. Section 3 explores the in-storage implementation of S-W. Simulation results are discussed in Section 4. Section 5 presents a discussion on the scalability of a ReCAM PRinS system, and Section 6 offers conclusions.

## 2 ReCAM-Based Processing-in-Storage

Resistive memories store information by modulating the resistance of nanoscale storage elements. They are nonvolatile, free of leakage power, and emerge as long-term potential alternatives to charge-based memories, including NAND flash. The metal-oxide resistive random access memory (ReRAM), employing one resistive device and possibly also one transistor (1R1T) per bit-cell, is considered a potential technology to replace next-generation nonvolatile memories. Its main features are high reliability and fast access speed. A test-chip of 32GB device with two ReRAM-based memory layers and a CMOS logic layer underneath has been developed [2], demonstrating design techniques to achieve a high density functional chip.

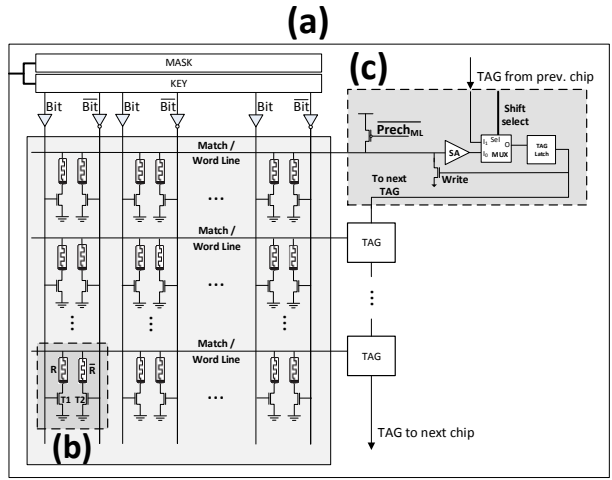

Figure 1. (a) single ReCAM crossbar IC. (b) 2T2R ReCAM bitcell. (c) TAG logic.

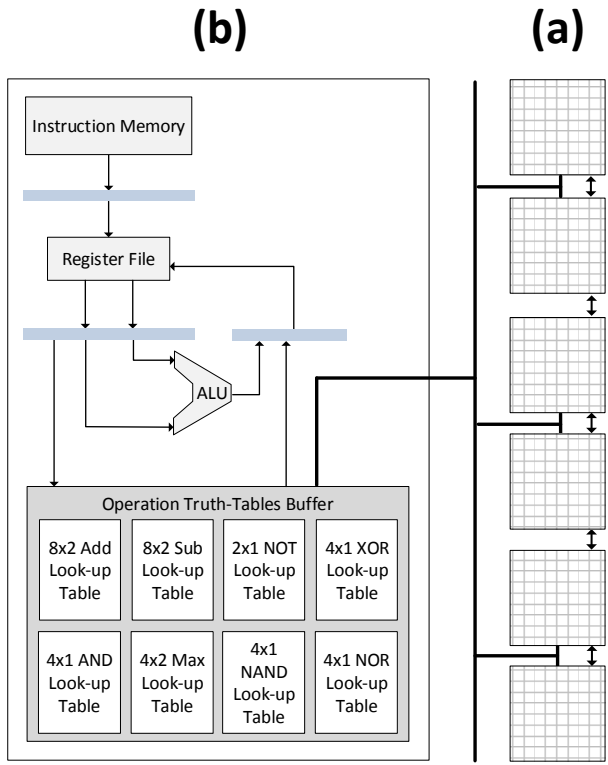

Figure 2. ReCAM-based Storage system is composed of separate multiple ICs (a). The ICs are connected by a reduction network to the microcontroller (b).

TABLE 1

OPERATIONS USED IN S-W SCORE CALCULATION

| Instruction | Cycles |
|---|---|
| **32 bit** | |
| Shift down one row | 96 |
| B ← A + B | 256 |
| C ← A + B | 512 |
| Row-wise Max (A, B) | 64 |
| Max Scalar (A) | 64 |
| **2 bit** | |
| DNA Base-Pair Match | 10 |

## 2.1 ReCAM Crossbar Array

While ReRAM may employ one transistor and one memristor (1T1R) cells, ReCAM uses 2T2R cells, following [3][4]. Figure 1(a) shows the resistive CAM crossbar. A bitcell, shown in Figure 1(b), consists of two transistors and two resistive elements (2T2R). The KEY register contains a data word to be written or compared against. The MASK register defines the active columns for write and read operations, enabling bit selectivity. The TAG register (Figure 1(c)) marks the rows that are matched by the compare operation and may be affected by a parallel write. The TAG register enables chaining multiple ReCAM ICs.

In a conventional CAM, compare operation is typically followed by a read of the matched data word. When in-storage processing involves arithmetic operations, a compare is usually followed by a parallel write into the unmasked bits of all tagged rows, and additional capabilities, such as read and reduction operations, are included [5].

Any computational expression can be efficiently implemented in ReCAM storage using line-by-line execution of the truth table of the expression [5]. Arithmetic operations are typically performed bit-serially. Table 1 lists the operations used in S-W implementation (Section 3) and the number of cycles required per each one. Shifting down a consecutive block of rows by one row position requires three cycles per bit. First, compare-to-'1' copies the source bit-column of all rows into the TAG. Second, shift moves the TAG vector down by setting the shift-select line (Figure 1(c)). Third, write-'1' copies the shifted TAG to the same bit-column. Shifting 32-bit numbers thus requires 96 cycles. Addition (in-place or not) is performed in a bit-serial manner us-ing a truth table approach [5] (32 bits times 8 truth-table rows times 2 for compare and write amount to 512 cycles). Row-wise maximum compares in parallel two 32-bit numbers in each row. Max Scalar tags all rows that contain the maximal value in the selected element. Additional operations, such as parallel and reduction arithmetic, may be required for other algorithms.

## 2.2 ReCAM PRinS System Architecture

Conceptually, the ReCAM comprises hundreds of millions of rows, each serving as a computational unit. The entire array may be divided into multiple smaller ICs (due to power per die restrictions, Figure 2(a)), which use the same MASK and KEY. A row is fully contained within an IC. All ICs are daisy chained for Shift and Max Scalar operations. Therefore, in practice, operations listed in Table 1 take several more cycles to enable inter-IC shift operations.

The ReCAM processing-in-storage system uses a microcontroller (Figure 2(b)) similar to [6]. It issues instructions, sets the key and mask registers, handles control sequences and executes read requests. In addition, the microcontroller holds the associative instructions buffer, containing the truth tables for associative instructions. Since instructions are performed bit-serially, these tables are typically small, as evident in Figure 2(b). Part of the associative instructions buffer is user-programmable with custom instructions, such as the match operation in Table 1.

## 3 ReCAM PRinS Smith-Waterman Implementation

### 3.1 Smith-Waterman Algorithm

S-W identifies the optimal local alignment of two sequences by computing a two-dimensional scoring matrix $H$ [7]. Each $H_{i,j}$ element is calculated according to Eq. (3). $\sigma(a_i, b_j)$ is the match score between the base-pairs in row $i$ ($i^{th}$ element of sequence A) and column $j$ ($j^{th}$ element of sequence B). Matching base-pairs score positively (e.g., +2), while mismatching result in negative score (e.g., -1). The optimal alignment score between two sequences is the highest score in the matrix $H$.

$$E_{i,j} = \max\{E_{i,j-1} - G_{ext}\ (i)\ ;\ H_{i,j-1} - G_{first}\ (ii)\} \quad (1)$$

$$F_{i,j} = \max\{F_{i-1,j} - G_{ext}\ (i)\ ;\ H_{i-1,j} - G_{first}\ (ii)\} \quad (2)$$

$$H_{i,j} = \max\begin{cases} H_{i-1,j-1} + \sigma(a_i, b_j) & (i); \\ E_{i,j}\ (ii);\ F_{i,j}\ (iii); 0\ (iv) \end{cases} \quad (3)$$

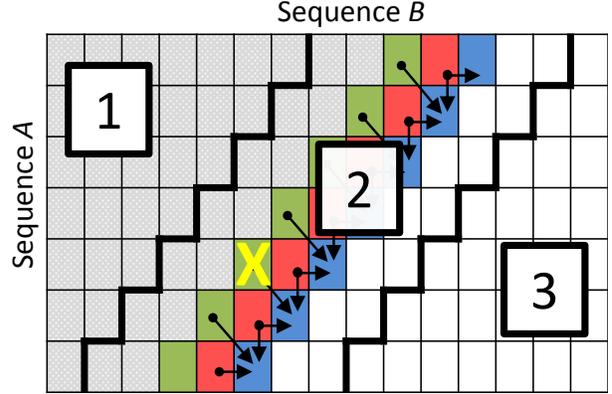

Figure 3. Parallel S-W scoring snapshot of matrix $H$. Thick borders separate ReCAM implementation to three logical sections (Section 3.2). The greyed-out cell scores have already been computed. The three plain-colored antidiagonals are stored in ReCAM. The green and red antidiagonals are used to compute the score of the blue antidiagonal. White-colored cell scores are yet to be computed. The cell marked with X contains the global maximum score.

The alignment may contain gaps in both sequences which are penalized in the score calculation (by negative scores). According to the affine gap model [8], opening a gap is harder than extending it, therefore the penalty for opening a gap is larger. The affine penalty scheme is calculated with two additional matrices, $E$ and $F$, equations (1) and (2); $G_{first}$ and $G_{ext}$ are the penalties for starting and extending a gap, respectively. The matrices $E$, $F$ and $H$ are initialized with $E_{0,j} = E_{i,0} = F_{0,j} = F_{i,0} = H_{0,j} = H_{i,0} = 0$ for all $i$ and $j$.

Filling the scoring matrix $H$ is the computationally intensive part of S-W. In a sequential implementation of the algorithm, cell filling is performed in either row- or column-wise order. A parallel implementation allows all independent cells to be computed in the same iteration. Such cells reside on the same antidiagonal. The matrix is filled along the main diagonal, as illustrated in Figure 3.

The sequential time complexity is $O(nm)$, where $n$ and $m$ are the respective lengths of the sequences. Parallel time complexity on $p$ parallel processing units is $O(nm/p)$. In ReCAM, the processing unit is a memory row. Since ReCAM may comprise hundreds of millions of rows, unlike GPU or FPGA implementations, $p$ could possibly be larger than $max\{n, m\}$ even for very large $n$ and $m$. Hence, ReCAM can achieve linear time complexity of $O(max\{n, m\})$.

## 3.2 Smith-Waterman Algorithm in ReCAM PRinS

In this work we focus on finding the maximal alignment score. Therefore storing the entire matrix in memory is not needed. This is in contrast to the full algorithm which also contains the traceback part for finding the alignment [7].

A total of four antidiagonals is required to compute a new antidiagonal of $H$: two of $H$ (see Eq. (1)-(3) and green, red of Figure 3), one of $E$ (see Eq. (1)) and one of $F$ (see Eq. (2)). Thus, five matrix antidiagonals are stored in the ReCAM in each iteration (E, F, AD[0], AD[1] and AD[2] in Figure 3 and Figure 5). A *tmp* field stores partial results. The overall space complexity required for executing the algorithm is therefore $O(\min\{n,m\})$.

Each of the five antidiagonals is mapped onto a 32-bit column in the ReCAM. Every ReCAM row retains one element of the vectors *seqA, SeqB, E, F, AD[0], AD[1], AD[2]* and *tmp*. The first two numbers are the 2-bit elements of sequences A and B, respectively.

S-W algorithm implementation on ReCAM can be divided into three logical sections. The first section, marked 1 in Figure 3, starts at the top-left cell and covers a triangle with each edge of length min{m,n} cells. In it, the most recently scored antidiagonal is longer by one cell than the previous one. The third section (3 in Figure 3) is of a similar shape and same dimensions, ending at the bottom-right cell. In it, every new scored antidiagonal is one cell shorter than the previous one. The second section (2 in Figure 3) is a parallelogram between the first and third sections. In it, all antidiagonals are of the same length.

Figure 4 presents the pseudocode of the S-W score finding on ReCAM. Three ReCAM columns are required to store last two scored antidiagonals of $H$ and the presently computed one, notated as *AD[2]-AD[0]* in code. During execution, these columns are cyclically buffered; the oldest scores are replaced by the new ones (line 4 in Figure 4). Additional three 32-bit columns are used to store antidiagonals of *E, F* and *tmp*.

Figure 5 shows a ReCAM crossbar snapshot at the beginning (a) and the end (b) of a single iteration of Figure 4. At line 5, *seqB* is shifted one ReCAM row down in order for all to-be-matched base-pairs to reside in the same ReCAM rows (second column from left in Figure 5). *AD[left_AD]* is also shifted one row down for the matching cells to be aligned (line 6 in Figure 4, and *AD[0]* in Figure 5). After calculating the

```
SmithWatermanScore(A, n, B, m) {
1   init (tmp, AD[2…0][*], F[*], E[*], seqA[*], seqB[*]) ← (0,…,0,A,0)
2   max_score ← 0   //scalar to hold the maximal cell value
3   for i=0 to n+m-1 do {
4     right_AD ← i mod 3; middle_AD ← (i – 1) mod 3; left_AD ← (i – 2) mod 3
5     seqB[*] ← B[i…1]  // Prepare subsequence B for next iteration
6     shift AD[left_AD][*]  1 row down
7     AD[right_AD][*] ← AD[left_AD][*] + match(seqA[*], seqB[*])  // (i) in Eq. (3)
      //AD[left_AD] is not needed anymore. Will be used as a temp variable
8     AD[right_AD] ← max{ AD[right_AD][*], 0}        // (iv) in (3)
9     AD[left_AD][*] ← AD[middle_AD][*] – $G_{first}$  // (ii) in (1) & (2)
10    tmp ← F[*] – $G_{ext}$
11    F[*] ← max{ AD[left_AD][*], temp }             // (i) in (2)
12    AD[right_AD][*] ← max{ AD[right_AD][*], F[*]}  // (ii) in (3)
13    temp ← E[*] – $G_{ext}$
14    E[*] ← max{ AD[left_AD][*], tmp }              // (i) in (1)
15    shift E[*]  1 row down
16    AD[right_AD][*] ← max{ AD[right_AD][*], E [*]}  // (iii) in (3)
17    max_score ← max{maxScalar(AD[right_AD][*]), max_score }  //scalar inst.
   }
```

Figure 4. Pseudo-code of S-W algorithm on ReCAM

matching score (line 7), *AD[left_AD]* is no longer required and is therefore used to store temporary results. Next, the max between the match score and zero is calculated (line 8). Note that (ii) in equations (1) and (2) belong to the same antidiagonal, therefore it is enough to calculate (ii) once for both *E* and *F* (line 9). Lines 10-16 compute equations (1)-(3). In line 15, after *E* is calculated, its columns are shifted one row down to have the values of *E* aligned with the appropriate ones in *AD[right_AD]*. In section 1 and 2 of Figure 3, the down-shifted columns require zero-padding of the top-most ReCAM row (not shown in Figure 4). At the end (line 17), the global max is updated with the maximal *H* cell score. After a specific base-pair of *seqB* has been aligned with all *seqA* base-pairs, it is cyclically shifted to its original position (not displayed in Figure).

The total number of iterations is the sum of lengths of the two sequences. Each iteration performs 17 instructions. The number of ReCAM rows affected by an instruction is marked by [*]. That number increases (decreases) in section 1 (3) and remains constant in section 2 (the minimum of the lengths of the two sequences).

At the beginning of execution (first cell of section 1), only the top-most ReCAM row is active. Each subsequent iteration activates an additional row until reach-ing min$\{m, n\}$ active rows in an iteration. During section 2, the number of active rows remains constant. During section 3, the number of active rows decreases, starting with the top-most row to be inactive and subtracting one active row in each iteration. The average number of active rows is $n \cdot m/(n + m)$.

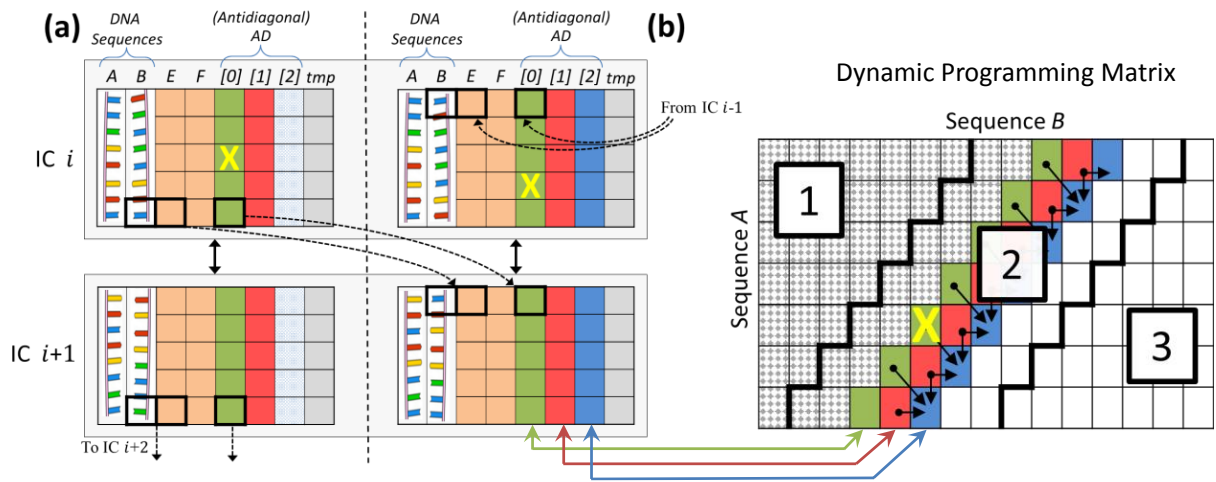

Figure 5. Organization of data in the ReCAM crossbar array at the beginning (a) and the end (b) of a single iteration of Figure 4. *AD[2]* contents is being replaced with the new result. Bottom rows in a crossbar IC are daisy-chained to the next IC in a shift instruction.

## 4 SIMULATION

The S-W algorithm is simulated on ReCAM using the cycle-accurate simulator introduced in [5], employing ReCAM performance and power figures obtained by SPICE simulations. The simulated ReCAM parameters are listed in Table 2. Power figure was taken from [5].

The simulation employs sequence data retrieved from the National Center for Biotechnology Information (NCBI), comparing human (GRCh37) and chimpanzee (panTro4) homologous chromosomes, similar to [9]. The CUPS metric (Cell Updates per Second) is used to measure S-W performance. Results are compared to other works in Table 3. A multi-GPU implementation [9] reached 11.1 TCUPS on a cluster of 128 compute nodes with a total of 384 Tesla M2090 GPUs. An FPGA implementation of S-W reaches 6.0 TCUPS on the RIVYERA platform [10] having 128 Xilinx Spartan-6 LX150 FPGAs. A four Xeon Phi implementation achieves 0.23 TCUPS [11]. On ReCAM, we demonstrate 53 TCUPS, computing a total of $57.2 \times 10^{12}$ scores. The table also shows computed GCUPS/Watt ratios; ReCAM is close to twice better than the FPGA solution and 80× better than the GPU system.

TABLE 2
SIMULATED ReCAM PARAMETERS

| ReCAM Parameter | Value |
| --- | --- |
| Active storage size | 8GB |
| Frequency | 1Ghz |
| Power per Integrated Circuit | 200W |
| Number of Integrated Circuits | 32 |

TABLE 3
SUMMARY OF STATE-OF-THE-ART PERFORMANCE FOR S-W SCORING STEP IN PREVIOUS WORKS AND IN ReCAM

| Accelerator | Xeon Phi | FPGA | GPU | ReCAM |
| --- | --- | --- | --- | --- |
| Perf. (TCUPS) | 0.23 | 6.0 | 11.1 | 53 |
| Number of ICs | 4 | 128 | 384 | 32 |
| Power (kWatt) | 0.8 | 1.3 | 100.0 | 6.6 |
| GCUPS/Watt | 0.3 | 4.7 | 0.1 | 8.0 |
| Reference | [11] | [10] | [9] | |

TABLE 4
ReCAM AND MULTI-GPU [9] PERFORMANCE

| Chr. | Table Size ($10^{12}$ Cells) | Max. Perf. of [9] (TCUPS) | ReCAM Perf. (TCUPS) |
|---|---|---|---|
| chr1 | 57.2 | - | 53 |
| chr5 | 33.5 | 11.1 | 41.8 |
| chr8 | 21.1 | 10.4 | 30.8 |
| chr16 | 8.1 | 9.7 | 19.3 |

The simulated ReCAM PRinS power dissipation is 6.6kW. The optimal setting to sustain this power figure with minimal performance overhead is dividing the ReCAM into 32 separate ICs, each with 256MB and 8M rows. The multi-GPU implementation using 384 Tesla M2090 GPUs and 256 Intel Xeon E5-2670 CPUs might dissipate 100kW, 15× higher power. Table 4 shows additional comparisons of ReCAM and the multi-GPU cluster [9], demonstrating up to 3.7× faster execution and 4.7× higher throughput on ReCAM.

## 5 SCALABILITY OF ReCAM PRinS SEQUENCE ALIGNMENT

Consider the case of one billion organism sequences. Each sequence is hundreds of millions base-pairs in size on average. Analyzing the contents of these sequences can lead to discoveries such as identification of disease-carrying genes, determination of evolutionary events and identification of regions that can be used to silence genes [12]. Performing an all-to-all alignment of the entire sequence database in a conventional data-center is not scalable. Every two sequences will require fetching to the main memory, close to the processing unit (CPU or accelerator). The high communication cost between separate storage units causes the system to be I/O bound in an all-to-all type of computation.

On the other hand, ReCAM-based storage is more scalable. Its inherent parallelism allows for scalability when adding more ICs, increasing storage capacity at no performance cost. The compute capability is linearly scalable in the number of ICs. Therefore, performing an all-to-all alignment of large sets, such as one billion sequences, does not require external communication for the ReCAM, in contrast to datacenter-scale storage. A more effective solution, in terms of performance and energy, is using ReCAM as primary storage when large alignment operations are constantly performed.

## 6 CONCLUSIONS

This paper explores PRinS (Processing-in-Storage) implementation for the scoring step of the Smith-Waterman DNA sequence alignment algorithm on a novel solid-state storage device, based on Resistive Content Addressable Memory (ReCAM). ReCAM enables storage with *in-situ* processing capabilities. It can contain hundreds of millions of data rows, each serving as a processing unit. The proposed ReCAM PRinS system is divided into multiple ICs to accommodate power density constraints.

The sheer number of database searches on whole genomes creates a need for considerably higher performance than exists today. For example, aligning two very long sequences, such as complete human and chimpanzee chromosomes, is a difficult task for contemporary accelerators. Since the ReCAM PRinS contains hundreds of millions of PUs, its performance increases with input size. We show that ReCAM PRinS has the potential to provide 4.7× performance improvement and 15× lower power dissipation over a 384-GPU cluster.

This research can be extended in several ways: First, the ReCAM PRinS S-W scoring algorithm can be extended to provide complete DNA sequence alignment (i.e., both matrix-fill and traceback steps), maintaining the same performance and power advantages. Second, the ReCAM PRinS algorithm can be applied in parallel to complete DNA sequences of two organisms, and not only to specific chromosomes. Third, the proposed S-W ReCAM PRinS algorithm can be applied to the wider challenge of aligning protein sequences. That problem is more challenging than DNA alignment because the required substitution matrix is typically 20×20 rather than 2×2, and the ReCAM could store the entire substitution matrix, resulting in efficient parallel processing.

ReCAM PRinS architecture, capable of general purpose associative processing, can also be applied to other challenging problems, such as machine learning and graph algorithms.

## ACKNOWLEDGMENT

Present work was partially funded by the Intel Collaborative Research Institute for Computational Intelligence.

## SHORT AUTHOR BIOS

**Roman Kaplan** received his BSC and MSc from the faculty of Electrical Engineering, Technion, Israel in 2009 and 2015, respectively. He is now a PhD candidate in the same faculty under the supervision of Prof. Ran Ginosar.
*Email: Romankap@gmail.com*

**Leonid Yavits** received his MSc and PhD in Electrical Engineering from the Technion. After graduating, he co-founded VisionTech where he co-designed a single chip MPEG2 codec. Following VisionTech's acquisition by Broadcom, he co-founded Horizon Semiconductors where he co-designed a Set Top Box on chip for cable and satellite TV. Leonid is a postdoc fellow in Electrical Engineering in the Technion. He co-authored a number of patents and research papers on SoC and ASIC. His research interests include non von Neumann computer architectures and processing in memory.
*Email: leonid.yavits@nububbles.com*

**Uri Weiser** is a Professor emeritus at the Electrical Engineering department, the Technion IIT and is in the advisory board of numerous startups.
He received the bachelor and master degrees in EE from the Technion and Ph.D in CS from the University of Utah, Salt Lake City.
Professor Weiser worked at Intel from 1988 till 2007. At Intel, Weiser initiated the definition of the Pentium® processor, drove the definition of Intel's MMX™ technology, invented the Trace Cache, co-managed the a new Intel Microprocessor Design Center at Austin, Texas and formed an Advanced Media applications research activity.
Professor Weiser was appointed an Intel Fellow in 1996, in 2002 he became an IEEE Fellow and in 2005 an ACM Fellow. In 2016 Professor Weiser received the IEEE/ACM Eckert-Mauchly Award for "leadership, as well as pioneering industry and academic work in high performance processors and multimedia architectures". The Eckert-Mauchly award is known as the computer architecture community's most prestigious award.
Prior to his career at Intel, Professor Weiser worked for the Israeli Department of Defense as a research and system engineer and later with National Semiconductor Design Center in Israel, where he led the design of the NS32532 microprocessor.
Professor Weiser was an Associate Editor of IEEE Micro Magazine and was Associate Editor of Computer Architecture Letters.
*Email: uri.weiser@ee.technion.ac.il*

**Ran Ginosar** received his BSc from the Technion—Israel Institute of Technology in 1978 (summa cum

laude) and his PhD from Princeton University, USA, in 1982, both in Electrical and Computer Engineering. His Ph.D. research focused on shared-memory multiprocessors. He worked at AT&T Bell Laboratories in 1982-1983, and joined the Technion faculty in 1983. He was a visiting Associate Professor with the University of Utah in 1989-1990, and a visiting faculty with Intel Research Labs in 1997-1999. He is a Professor at the Department of Electrical Engineering and serves as Head of the VLSI Systems Research Center at the Technion. His research interests include VLSI architecture, manycore computers, asynchronous logic and synchronization, networks on chip and biologic implant chips. He has co-founded several companies in various areas of VLSI systems.
*Email: uri.weiser@ee.technion.ac.il*